\documentclass[sigconf]{acmart}
\usepackage{makecell}
\usepackage{microtype}
\usepackage{tablefootnote}

\AtBeginDocument{%
  \providecommand\BibTeX{{%
    \normalfont B\kern-0.5em{\scshape i\kern-0.25em b}\kern-0.8em\TeX}}}

\acmConference[To Appear at HT '24]{To Appear at 35th ACM Conference on Hypertext and Social Media}{September 10--13, 2024}{Poznan, Poland}
\acmBooktitle{To Appear at 35th ACM Conference on Hypertext and Social Media (HT '24), September 10--13, 2024, Poznan, Poland}
\begin{document}

\title{Constructing a Common Ground: Analyzing the quality and usage of International Auxiliary Languages in Wikipedia}
\author{Marta Alet}
\email{m.aletpuig@gmail.com}
\affiliation{%
  \institution{Pompeu Fabra University}
  \city{Barcelona}
  \state{Catalonia}
  \country{Spain}
}
\authornote{This work was done during this author's master studies, and it's not related with her current position at Amazon.}
\author{Diego Sáez-Trumper}
\email{diego.saez@upf.edu}
\affiliation{%
  \institution{Pompeu Fabra University}
  \city{Barcelona}
  \state{Catalonia}
  \country{Spain}
}

\begin{abstract}
International Auxiliary Languages (IALs) are constructed languages designed to facilitate communication among speakers of different native languages while fostering equality, efficiency, and cross-cultural understanding. This study focuses on analyzing the editions of IALs on Wikipedia, including Simple English, Esperanto, Ido, Interlingua, Volapük, Interlingue, and Novial. We compare them with three natural languages: English, Spanish, and Catalan. Our aim is to establish a basis for the use of IALs in Wikipedia as well as showcase a new methodology for categorizing wikis. We found in total there are 1.3 million articles written in these languages and they gather 15.6 million monthly views. Although this is not a negligible amount of content, in comparison with large natural language projects there is still a big room for improvement. We concluded that IAL editions on Wikipedia are similar to other projects, behaving proportionally to their communities' size. Therefore, the key to their growth is augmenting the amount and quality of the content offered in these languages. To that end, we offer a set of statistics to understand and improve these projects, and we developed a web page that displays our findings to foster knowledge sharing and facilitate the expansion of the IAL communities.
\end{abstract}




\keywords{Wikipedia, International Auxiliary Language, Inclusive web, Content Quality, Cross-cultural understanding, Knowledge sharing.}


\maketitle

\section{Introduction}
Wikipedia is ``the free encyclopedia that anyone can edit"~\cite{wiki:wikipedia}. Due to its availability in numerous languages, Wikipedia is accessible to users around the world. Similarly, International Auxiliary Languages (IALs) help us communicate with people with whom we do not share a common language. They allow two people with different first languages to understand the same text without favouring either of them. Because of their sociopolitical neutrality and simplicity, they aimed to bring together a separate world~\cite{zorrilla2018still}. IALs are designed to be easily learned and accessible to speakers of diverse languages. But ``language users make a language come alive", the challenge IALs face is that they lack a ``cultural base", and adoption is still relatively low~\cite{ke2015global}. It can be difficult to attract readership if there is a lack of content written in those languages, and even more difficult if that content is not attractive to readers~\cite{ke2015global}. This difficulty is reflected in Bitereyst's 1992 research~\cite{biltereyst1992language}, which highlights the challenges of distributing and popularizing content across linguistic and cultural barriers. Their work’s insights into the obstacles faced by literary works in small European countries highlight a similar predicament for IALs on Wikipedia: the need for a strategic enrichment of content that transcends these barriers.

By analyzing the IAL Wikipedia community, we can detect what the main interests are and where the focus should be going forward. We studied the following seven IALs: Simple English, Esperanto, Ido, Interlingue, Interlingua, Novial and Volapük. Table \ref{summaryIALs} shows a short description of each IAL. Moreover, we use the English, Spanish, and Catalan language editions for comparison purposes, as they are well-established editions of different sizes. This way we can discover if IALs are different from other editions or if it is the size of the wikis the main difference. Thus, our goal is to understand these language editions and provide insights into what is keeping them from reaching a higher audience. Also by analyzing the current interests of their readers, we can help authors provide more relevant articles and increase the popularity of this community.

\begin{table}
\caption{Short description of each IAL based on the English Wikipedia}
\label{summaryIALs}
\begin{tabular}{lp{6cm}}
\hline
Simple English & Controlled language based on English, which only contains a small number of words.\\ 
Esperanto  & Intended as a universal second language for international communication, its vocabulary, syntax, and semantics mainly derive from Indo-European languages, with about 80\% from Romance languages, and elements from Germanic, Greek, and Slavic languages. \\ 
Ido & Constructed language derived from a reformed version of Esperanto. \\ 
Interlingua & Derived from natural languages, it focuses on common vocabulary and grammar shared by Western European languages, primarily influenced by Latin and Greek. \\ 
Volapük & The grammar is based on European languages, while the vocabulary is mainly from English, with some German and French influences. \\ 
Interlingue & Aims for grammatical regularity and natural character with a vocabulary derived from Romance and Germanic languages, using recognized prefixes and suffixes. \\ 
Novial & Naturalistic IAL based largely on the Germanic and Romance languages while its grammar is influenced by English.\\
\hline
\end{tabular}
\end{table}

There is not much academic research about IALs on Wikipedia. In 2018 the “An Overview of International Auxiliary Languages in Wikipedia” study was uploaded to the internet~\cite{Ever_A_Garcia_M}. Their main findings were that Simple English was by far the most popular IAL; that there is indeed a community of editors of these languages, as they found a high percentage of co-editions in the articles; and the fact that authors who edit articles in one IAL often do so in another, meaning that it is a relatively unified community rather than
each auxiliary language having its own separate community, as is often the case with natural languages. However, there is still much to learn and study about this community, and the focus of this study is to deepen our understanding of the IAL community on Wikipedia as well as set a precedent for future research on the study of this community. We have created a website to showcase our findings and some useful tools we developed to help IAL's editors learn more about their projects\footnote{https://ial-tools.streamlit.app/}.

\section{Datasets}
\label{referenceAPIs}
We have gathered some basic statistics on each Wikipedia language edition in Table \ref{Table_Details}. The table is ordered by the
amount of monthly views, and going forward this will also be the order used in this paper.

For conducting our study we assembled two datasets. Both are composed of the content in Wikipedia from the seven IALs and the three natural languages stated.
\begin{table*}
  \caption{Details of each Wikipedia language edition (Monthly Average 2022)\cite{StatsWikimedia}}
  \label{Table_Details}
  \begin{tabular}{ccccccccc}
    \toprule
    \shortstack{Language}&\shortstack{Year of \\ creation} &\shortstack{\#Articles}&\shortstack{\#Editors}&\shortstack{\#Active Editors}&\shortstack{\#Monthly views}&\shortstack{\#Monthly user \\ views}&\shortstack{\#Articles created \\ in 2022}&\shortstack{\#Countries\\ with viewers}\protect\footnotemark[5]\\
    \midrule
    English & 2001 & 44.4M & 360.5K & 38.2K & 9.6B & 7.2B & 142518 & 251\\
    Spanish & 2001 & 5.8M & 68.9K & 4.5K & 1.1B & 842.4M & 76414 & 231\\
    Catalan & 2001 & 1.3M & 2.9K  & 445.7 & 42.9M & 13.1M & 26964 & 158\\
    \midrule
    Simple English & 2001 & 630.0K & 3.6K & 280.7 & 27.1M & 12.6M & 23200 & 214\\
    Esperanto & 2001 & 517.6K & 507.1 & 94.1 & 11.1M & 1.4M & 1926 & 119\\
    Ido & 2004 & 47.9K & 61.6 & 10.1 & 2.2M & 468.1K & 4190 & 70\\
    Interlingua & 2002 & 38.6K & 40.2 & 6.8 & 1.9M & 382.0K & 3467 & 80\\
    Volapük & 2003 & 41.2K & 32.1 & 5.8 & 1.7M & 299.8K & 2819 & 57\\
    Interlingue & 2004 & 14.8K & 28.6 & 4.3 & 972.8K & 261.3K & 2645 & 54\\
    Novial & 2006 & 4.1K & 13.8 & 1.6 & 427.2K & 172.7K & 35 & 42\\
  \midrule
    Total IALs & - & 1.3M & 4.3K & 403.4 & 45.4M & 15.6M & 38282 & 214\\ 
  \bottomrule
\end{tabular}
\end{table*}

The first, \textit{“Exploratory Dataset”}, contains all articles in these ten Wikipedias up until ‘2023-01-16'. This data was obtained from Wikimedia APIs: the Language-agnostic Quality Prediction\footnote{\url{https://meta.wikimedia.org/wiki/Research:Prioritization_of_Wikipedia_Articles/Language-Agnostic_Quality}} and Topic Classification APIs~\cite{johnson2021language}, as well as Wikimedia’s REST API~\footnote{\url{https://www.mediawiki.org/wiki/REST_API}}. We also utilized the Python pypopulation\footnote{https://pypi.org/project/pypopulation/} library to gather population data and seamlessly integrate it with the geographic aspect of the traffic the pages receive. The dataset includes statistics and information about the article's content and pageviews, such as the characteristics of the articles (their length, the amount of media they contain, their references, etc.), \footnotetext[5]{We consider that a country has viewers if the amount of views is above 100 as that is the lowest amount registered by the REST API.}\setcounter{footnote}{5}the topics that the article is associated with~\footnote{The API returns a probability of the article belonging to each topic. For the sake of this study, we will consider that an article is related to a topic if the probability is at least 0.5.}, the number of views distributed by country in January of 2023. We use this dataset to study the completeness of articles in each of the IALs, the topics that each language covers, and to see how distributed the views are for each of these auxlangs.

Moreover, some of the discussion will deliberately focus on a subset called \textit{"Top 100 Articles"}. This subset pertains to the articles that ranked among the 100 most viewed in January of 2023 within their respective language editions. This examination is conducted with the aim of uncovering the reasons behind their popularity within these communities. The reason for selecting the top 100 articles, as opposed to any other quantity, is grounded in the observation that these articles represent the core of reader engagement and interest across the IAL communities on Wikipedia. This selection captures the most attractive content, where a significant portion of views are concentrated, providing a rich and manageable dataset for analysis. Extending this scope to include a larger set of articles would dilute the focus on high-impact content without proportionately enhancing the insights into the dynamics of content popularity and reader preferences.

Next, we built a \textit{“Views Dataset”}, that contains only articles created between ‘2022-01-01' and ‘2022-12-31'. This dataset was obtained by making SQL queries in Quarry, a tool that provides access to replicas of Wikimedia’s databases\footnote{https://quarry.wmflabs.org}. Additionally, we used it to collect the views of the first month after creation of all articles. The number of articles created in each language can be seen in Table \ref{Table_Details}. The goal behind this dataset is to analyze the content generated in a specific time range and track its evolution just after its creation. In order to make cross-language comparisons, we merged the data gathered using the QID\footnote{QID refers to the article's identifier on Wikidata, which is language agnostic.} of the articles.

The data used in this work can be found in this repository\footnote{\url{https://github.com/MartaAlet/IALs_wikipedia_dataset}}.

\section{Data Exploration}
\subsection{Topic Distribution in the IAL Community}
\begin{figure}
\caption{(left) Distribution of topics by language; (right) Distribution of Topics by Language of the Top 100 Articles }
\label{distributiontopics}
  \begin{minipage}{0.48\columnwidth}
    \includegraphics[width=\textwidth]{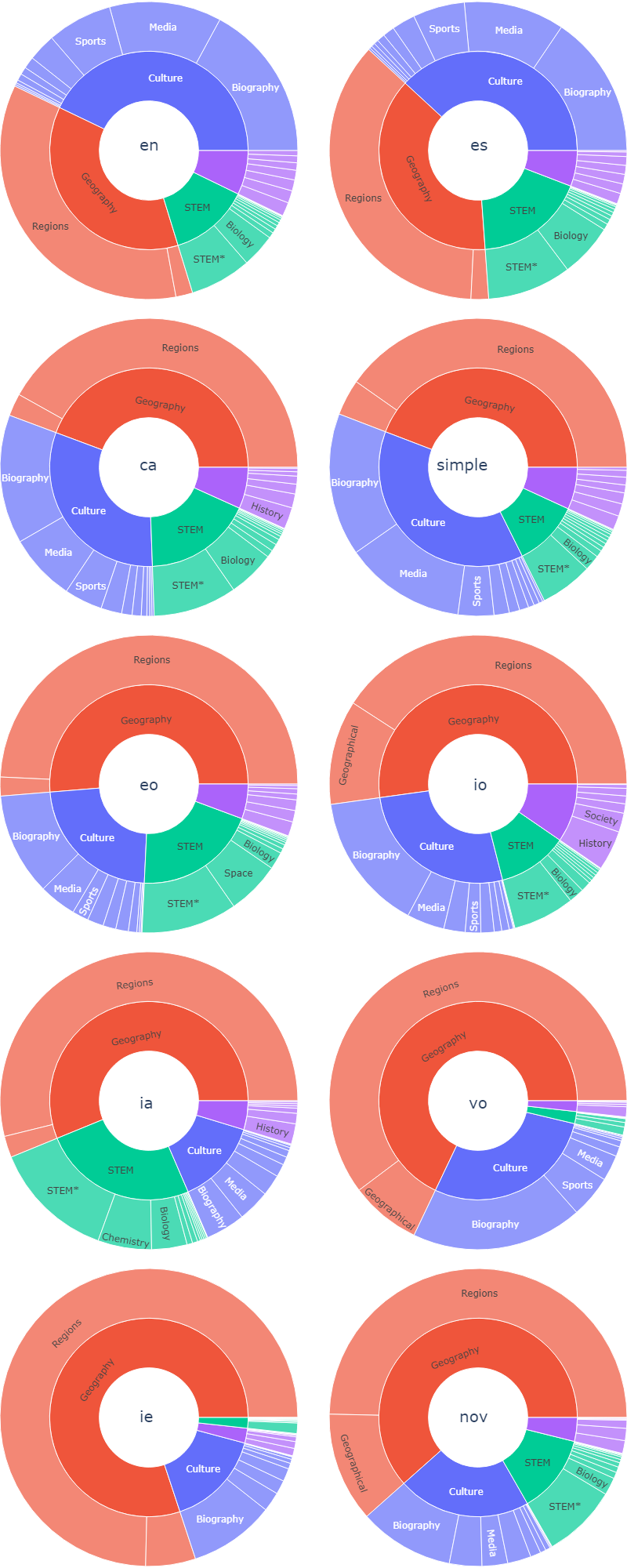}
  \end{minipage}
  \hfill
  \vline
  \hfill
  \begin{minipage}{0.48\columnwidth}
    \includegraphics[width=\textwidth]{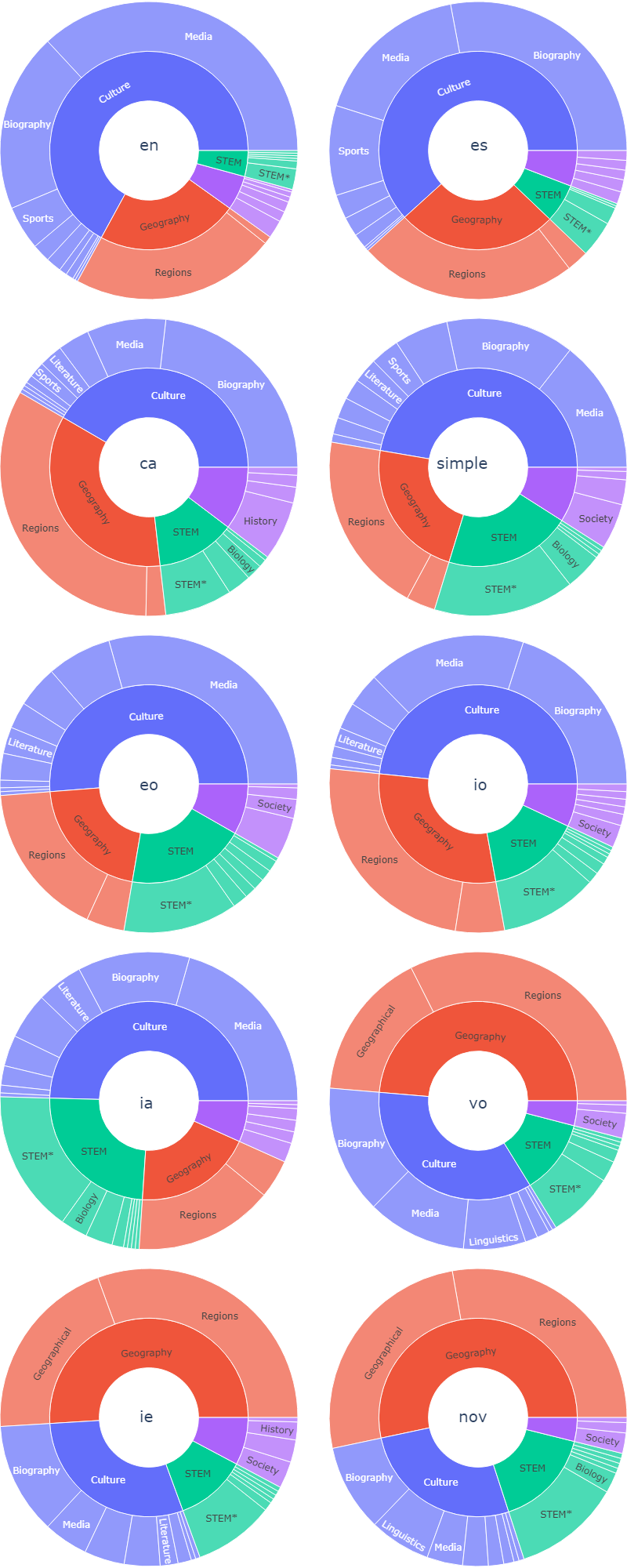}
  \end{minipage} 
  \vspace{1em}
  \newline
  \begin{minipage}{0.96\columnwidth}
    \includegraphics[width=\textwidth]{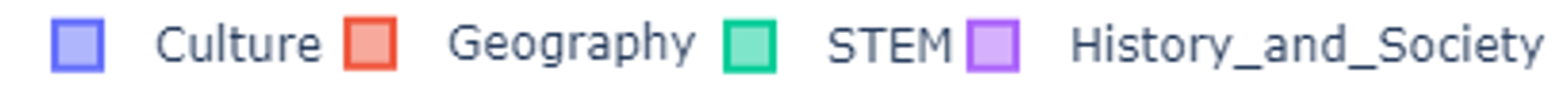}
  \end{minipage}
\end{figure}

Here we analyze the distribution of topics of the content available in these editions and assess if it aligns with the interests of the user base in these communities by comparing it to the most popular articles. We studied the topical distribution of each Wikipedia IAL edition, by using the Language-agnostic Topic Classification API ( Section \ref{referenceAPIs}). The API returns the main topic(s) of each article, following the ORES article's topic taxonomy\footnote{https://www.mediawiki.org/wiki/ORES/Articletopic}. We grouped the topics into their second-level theme, Figure \ref{distributiontopics} (left) shows the distribution of these topics by language. We can see that Geography (red) is the most frequent topic in all language editions. Its most popular subtopic is Regions, meaning we can easily assign a geographical region to the articles in all of the IALs. Given these results, we decided to analyze the Geographic topic separately (will be discussed in section \ref{sec:world-distribution}). We can see that the second most dominant topic is \textit{Culture-Biography} in all languages except Interlingua, which has a higher percentage of STEM articles. The main findings are that Interlingue’s topic distribution does not resemble any of the other languages. We can also observe that Catalan and Spanish have a very similar distribution of topics, which is to be expected given they have a common cultural background. The main takeaway is that the IAL community is diverse and each IAL has a different distribution on the percentage of articles they offer on a given topic.

By filtering the articles that were in the list of the 100 most viewed articles in January 2023, we get Figure \ref{distributiontopics} (right), where we show the topical distribution of the popular articles. We see that these distributions do not match the ones seen previously. Volapük, Interlingue, and Novial have somewhat similar distributions. Simple English and Esperanto also seem to have similar proportions.

Comparing both sides of Figure \ref{distributiontopics} we have seen that the content offered is not completely in line with the interests of their respective user-base. While we see a large number of articles related to geography in all editions, the users of all languages except those in Interlingue and Novial have a higher tendency to look at culture or history-related content.
\subsection{Quality Analysis of the Articles}
The characteristics of the article, such as page length\footnote{Article size in bytes}, number of references, wikilinks, categories, media, and headings, give us some insights into how intricate and developed the content offered in each language edition~\cite{articleq}. The mean value for all of these features for each one of the IALs can be viewed in Table \ref{features}. Simple English tends to have longer articles, higher numbers of references, and more categories compared to other IALs. Esperanto has a higher average number of headings, suggesting a structured organization of content within articles. Ido stands out with a higher average number of wikilinks and media in its articles. This indicates that the editors of the Ido language version may prioritize extensive cross-referencing and adding visual content. Considering that Novial has 4.1K articles and 13.8 average editors, it's notable that the statistics for page length, number of media, and number of headings are relatively high compared to the other IALs. This suggests that despite having a lower number of articles and editors, its editors might prioritize comprehensive content, multimedia inclusion, and structured organization within its articles slightly more than the editors in other IALs. The remaining languages (Interlingua, Volapük, and Interlingue) generally have lower values across the measured metrics, denoting a lower level of development.
\begin{table}
\setlength{\tabcolsep}{3pt}
\caption{Mean features of articles by IAL}
\label{features}
\begin{tabular}{lrrrrrrr}
\hline
\multicolumn{1}{c}{\textbf{Feature}} & \multicolumn{1}{c}{\textbf{simple}} & \multicolumn{1}{c}{\textbf{eo}} & 
\multicolumn{1}{c}{\textbf{io}} & 
\multicolumn{1}{c}{\textbf{ia}} & 
\multicolumn{1}{c}{\textbf{vo}} & 
\multicolumn{1}{c}{\textbf{ie}} & 
\multicolumn{1}{c}{\textbf{nov}} \\ \hline
Page length & 3.2K & 2.8K & 1.5K & 0.8K & 0.7K & 0.9K & 1.9K \\ 
\# references  & 3.0 & 1.0 & 0.4 & 0.4 & 0.1 & 0.1 & 0.3  \\ 
\# wikilinks & 25.3 & 26.4 & 17.8 & 7.7 & 5.2 & 9.0 & 19.9  \\ 
\# categories & 3.2 & 2.2 & 1.2 & 1.0 & 2.5 & 1.2 & 0.7 \\ 
\# media & 1.2 & 1.6 & 1.4 & 0.2 & 0.7 & 0.8 & 1.2 \\ 
\# headings & 2.1 & 2.8 & 1.2 & 0.8 & 0.7 & 0.9 & 2.1 \\ \hline
\end{tabular}
\end{table}

Considering the articles that were in their top 100 list in January 2023, we built Table \ref{qualityTop100}. The metrics of popular articles are higher than the average in all language editions. In comparison, the most viewed articles are 410\% longer, contain 1050\% more references, 287\% more wikilinks, 21\% more categories, 312\% more media, and 283\% more headings compared to the average articles. These statistics indicate that popular articles are more comprehensive, providing detailed information, with extensive referencing, and higher visual content. This suggests that the readers of the IAL community prefer more extensive and in-depth material, or that articles that attract more reader attention also get more editors' work. 
\begin{table}
\setlength{\tabcolsep}{3pt}
\caption{Mean features of top 100 articles by IAL}
\label{qualityTop100}
\begin{tabular}{lrrrrrrr}
\hline
\multicolumn{1}{c}{\textbf{Feature}} & \multicolumn{1}{c}{\textbf{simple}} & \multicolumn{1}{c}{\textbf{eo}} & 
\multicolumn{1}{c}{\textbf{io}} & 
\multicolumn{1}{c}{\textbf{ia}} & 
\multicolumn{1}{c}{\textbf{vo}} & 
\multicolumn{1}{c}{\textbf{ie}} & 
\multicolumn{1}{c}{\textbf{nov}} \\ \hline
Page length & 10.7K & 14.6K & 8.4K & 10.4K & 2.8K & 4.8K & 5.5K \\ 
\# references  & 11.6 & 11.1 & 4.5 & 16.1 & 0.8 & 1.2 & 3.3  \\ 
\# wikilinks & 76.2 & 145.3 & 127.0 & 80.8 & 14.9 & 35.0 & 41.9  \\ 
\# categories & 3.5 & 3.8 & 1.2 & 1.6 & 2.3 & 1.4 & 1.1 \\ 
\# media & 3.4 & 7.1 & 6.3 & 0.7 & 2.9 & 1.7 & 2.2 \\ 
\# headings & 5.9 & 9.2 & 6.1 & 4.5 & 2.8 & 3.8 & 2.4\\ \hline
\end{tabular}
\end{table}

These observations provide some insights into the quality of the articles of the IAL community. However, it's important to consider that article quality is a complex concept that cannot be fully captured by a few metrics alone. In any case, the careful edition with updated links, rich references and extensive explanation is a sure ground to build on a more attractive and engaging article.
\subsection{Geographical Distribution}
\label{sec:world-distribution}
\begin{figure}[!th]
 \caption{Monthly views per Wikipedia language and country divided by population}
\label{fig:mapaViews}
  \includegraphics[width = \columnwidth]{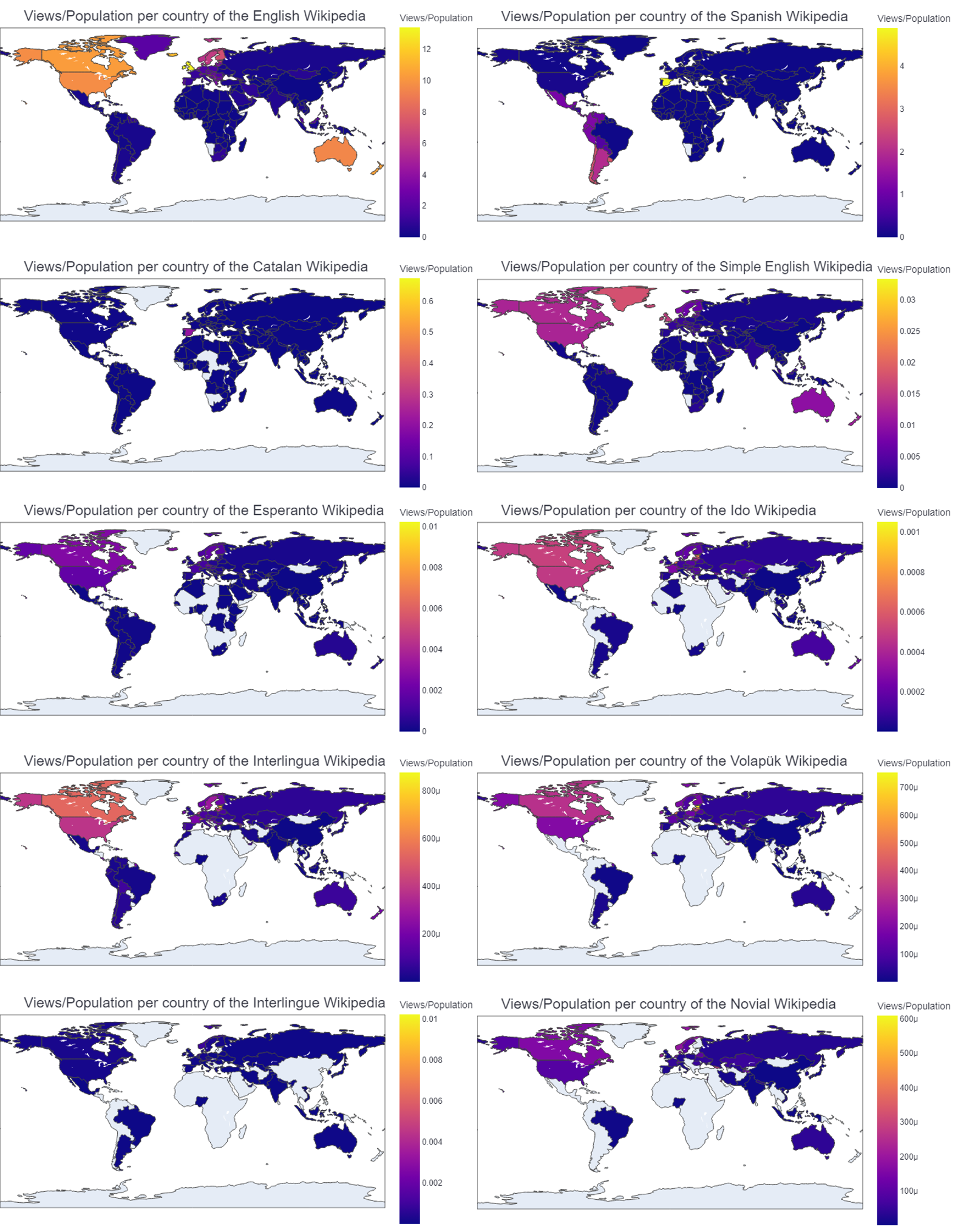}%
\end{figure}

We now turn our attention to the reach that the IAL editions have, both in terms of views per country as well as the number of countries. We also compare these distributions with the geographical aspect of the articles available in each edition. Figure \ref{fig:mapaViews} shows the distribution of monthly views of each wiki, according to the readers' location and the population of each country. We can see that English Wikipedia is the edition with a higher geographical reach and - not surprisingly - the one that has more viewers. Out of the IALs, Simple English is the most distributed in the number of countries where it is accessed. Spanish and Catalan are the only languages where the USA is not their main source of readers, it is Spain. Also in Spanish, we verify that countries of Latin America have a high concentration of readers. Notice that colours in Figure~\ref{fig:mapaViews} correspond to views of each country divided by its population. We should emphasize there is a correlation between the monthly traffic of each Wikipedia and the number of countries that the language edition reaches (see Table~\ref{Table_Details}). There is one outlier which is Simple English, that has almost the same distribution as English. Surprisingly, Interlingue seems to be evenly distributed globally, while other IALs usually concentrate their readers in a few countries. Lastly, we found that the viewers of Ido, Interlingua, and Volapük are concentrated in the Baltic region and North America.
Going deeper into this aspect of our research, we looked at what was the most popular IAL per country and found that in all countries it was Simple English and secondly Esperanto.

\begin{figure}
 \caption{Distribution of Geography Topics}
  \label{fig:topicsGeography}
\includegraphics[width = \columnwidth] 
  {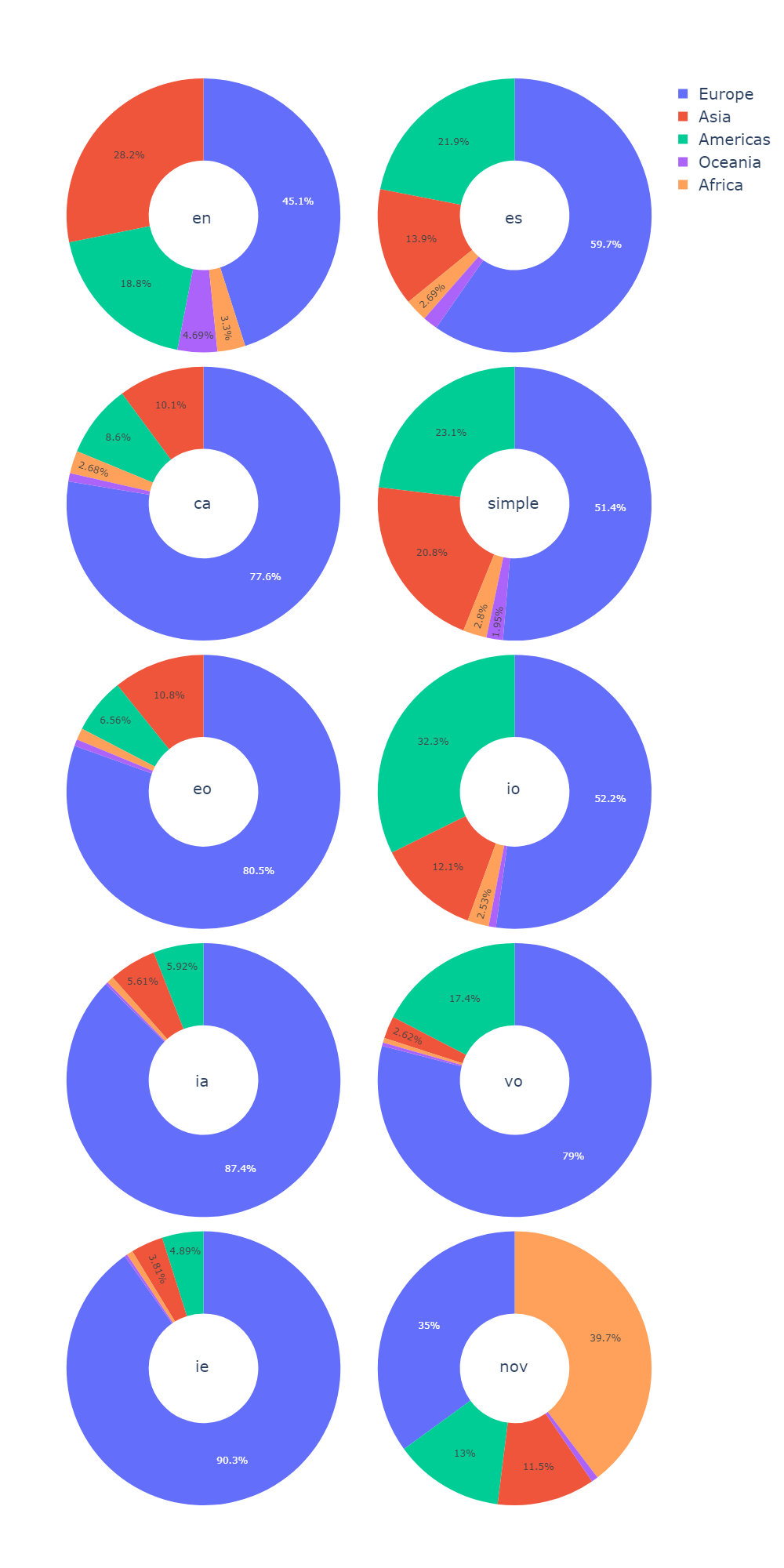}%
\end{figure}

While our previous analysis focused on the source of the readers, Figure \ref{fig:topicsGeography} is an analysis of the geographical aspect of the content of the articles. The majority of articles in all languages except Novial refer to Europe. Novial is the language with the evenest distribution of articles in terms of region, with Africa being the most frequently mentioned continent. Once again it can be seen that the contents of the IALs are not the same, however, we do notice similarities between some pairs: Interlingua and Interlingue, and Catalan and Esperanto, which covers similar regions. In general, we obtained a -0.49 Pearson correlation between the amount of articles in a language edition, and the percentage of content offered related to geography. We also found a -0.72 correlation between the average article length and this ratio.
\section{Viewership}
\subsection{Views just after article creation}
In this section, we analyze the viewership of the articles of each IAL during the first days after the article's creation and in comparison with their equivalents in other languages. For this task, we used the \textit{Views Dataset}, which contains only articles created in the ten wikis in 2022. Using the Wikidata items of the articles, we can relate them to other language versions and observe the correlation between their page views. The number of articles created in each language can be seen in Table  \ref{Table_Details}. Considering the fact that Novial had only 35 new articles in 2022, we decided to exclude this IAL from our study.
\subsection{Analysis of View Distribution}
\label{definitionPopularity}
We are interested in knowing how uneven is the distribution of views between popular and regular articles.  To measure this we apply the following procedure for each language:

\begin{itemize}
    \item Step 1: Order the articles by the total amount of views they received in the first month after their creation.
    \item Step 2: For each upper percentile we calculate the total sum of the first month's total views after the article's creation.
    \item Step 3: We compute the percentage that the total sum of views at the given popularity percentile represents.
\end{itemize}

\begin{figure}
\caption{Percentage of the total sum of views of all articles by upper percentile}
\label{fig:chart}
  \includegraphics[width = \columnwidth]{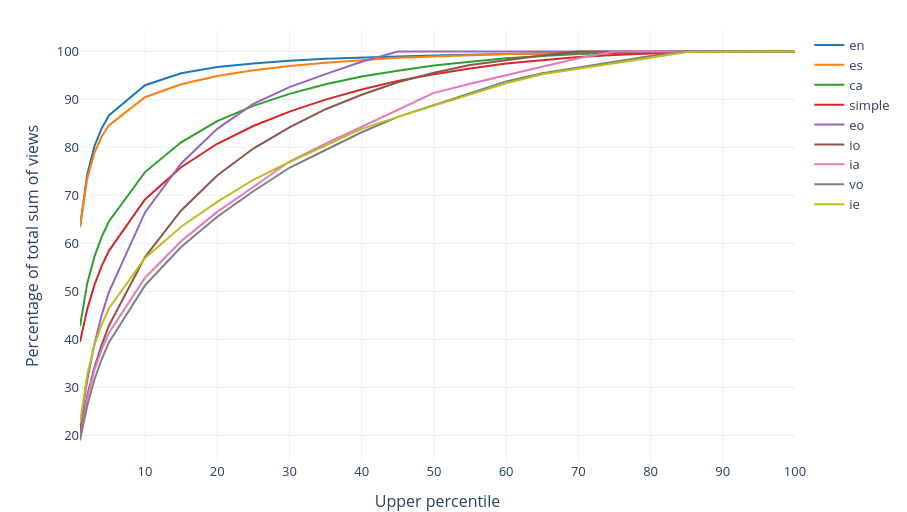}%
\end{figure}

Figure \ref{fig:chart} shows our results. As expected, all language editions follow a power law distribution. Notably, the curves are ordered according to the language edition's size: English and Spanish are the language editions that have the highest percentages of views attained by their top articles. Conversely, the smaller-sized IALs have lower curves, indicating that their most visited articles represent a comparatively lower percentage of the total traffic. Moreover, Catalan and Simple English show similar curves, which could be explained by their resemblance in size. We observe their 10th upper percentile has around 70\% of the total views of their dataset.


Having this in mind we could conclude that IALs can be considered aligned to the other language editions. The distribution of their views is similar to the established editions, and these similarities can be caused by their small size in comparison.
\subsection{Views according to Topic}
\begin{figure}
 \caption{Box plots depicting the distribution of the first-month views by Topic}
\label{fig:views_topics}
  \includegraphics[width=\columnwidth]{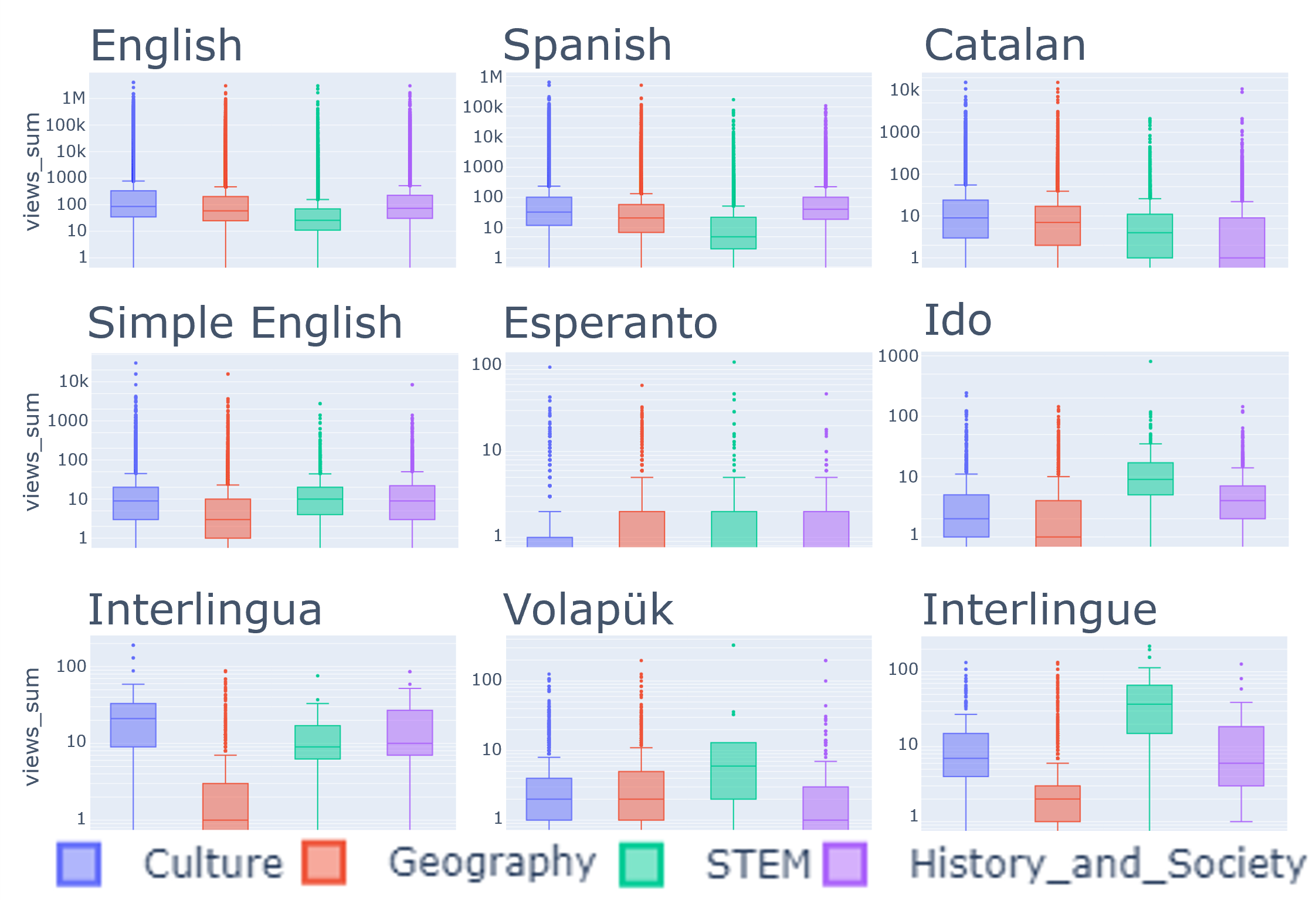}%
\end{figure}

As seen in previous sections, these wikis are not equally distributed when it comes to topics. It would pose the question if some topics generate more views than others in a specific wiki. For this section, we considered only the first level of categorizing: "Culture", "Geography", "History and Society", and "STEM". Figure~ \ref{fig:views_topics} confirms this idea, as we can see that each language edition has a specific topic that generates more views. In the case of Volapük, STEM articles generate 3 times more views than the other topics. Comparatively, Interlingua articles classified as Geography tend to have far fewer views than when the article relates to other topics. We also observe that the bigger the language edition is, the more balance there is between the topics in terms of views. These results also show the diversity of interests in each IAL. 
\section{Conclusions \& Future Work}
We've looked into the characteristics of the 1.3 million articles in the language editions of IALs and compared them to well-established Wikipedia editions (English, Spanish, and Catalan). We noticed that IALs are not very different from other language communities and that project size is the main factor that explains the usage of these projects. 

In order to address the current lack of content and low readership on IALs, we analyzed the factors that correlate with readership. For instance, we've been able to determine, that the articles which receive more views have higher quality. By focusing on content improvement and ensuring higher standards of quality, IAL communities could increase user engagement, encourage knowledge sharing, and facilitate community growth. Our study comparing readers and content, shed light on potential topics of interest for the IAL readers, that could help to grow these communities, attracting more readers and editors. We think there is potential for a tool such as a recommender system that could help expand the content available in these languages and likely enlarge the readership of these wikis. Suggesting missing pages across different languages to foster multilingual content growth was an idea presented recently and could very well help IALs grow on Wikipedia~\cite{wulczyn2016growing}. 

Also, this study has introduced a methodology for categorizing language editions on Wikipedia, showcasing a comprehensive approach to analyze and compare the popularity of articles, reader engagement, and topic distributions across different language editions. Although initially applied to International Auxiliary Languages, this methodology is adaptable and can be extended to a broader range of language groups. It offers valuable tools for both researchers and Wikipedia editors to identify key trends, content gaps, and opportunities for cross-linguistic content creation and enhancement.

To the best of our knowledge, this is the first large study on IALs on Wikipedia, and one of the few studies in this field about online content. There is still much to look into such as analyzing user behaviors, which could provide deeper insights into the functioning of the IAL community.

%
%
%
\section{Limitations}
The dataset used represents a snapshot of article views and topic distributions, which may evolve over time. Additionally, our study focuses on a specific set of IALs, and the findings may not be generalizable to other auxlangs or language communities. Future studies could expand the analysis to include additional IALs and investigate these topics further.




\begin{acks}
This work has been partially supported by the Department of Research and Universities of the Government of Catalonia (SGR 00930), and MCIN/AEI /10.13039/501100011033 under the Maria de Maeztu Units of Excellence Programme (CEX2021-001195-M).
\end{acks}

\bibliographystyle{ACM-Reference-Format}
\bibliography{sample-base}

\end{document}